\documentclass[mnextra]{mn}
\usepackage{amsmath, amssymb, epsfig}
\begin{document}
\def\etal{{et~al.}}
\def\mstar{{M$^{\star}_{b_{\rm J}}$}}
\def\Mpc{{$\,h^{-1}\,{\rm Mpc}$}}
\def\Mpcinv {{$\,h^{3}\,{\rm Mpc}^{-3}$}}
\def\apj {ApJ}
\def\apjl {ApJL}
\def\mnras {MNRAS}
\def\simlt{\mathrel{\rlap{\lower 3pt\hbox{$\sim$}} \raise
        2.0pt\hbox{$<$}}} \def\simgt{\mathrel{\rlap{\lower
        3pt\hbox{$\sim$}} \raise 2.0pt\hbox{$>$}}}
\def\simgt{\mathrel{\rlap{\lower 3pt\hbox{$\sim$}} \raise
        2.0pt\hbox{$>$}}} \def\simgt{\mathrel{\rlap{\lower
        3pt\hbox{$\sim$}} \raise 2.0pt\hbox{$>$}}}

\title[Luminosity and redshift dependent quasar clustering]
{Luminosity and redshift dependent quasar clustering}
\author[Cristiano Porciani \& Peder Norberg]
{Cristiano Porciani$^1$, Peder Norberg$^{1,2}$\\
$^1$ Institute for Astronomy, ETH Z\"urich, CH-8093 Z\"urich,
Switzerland \\
$^2$ SUPA\thanks{The Scottish Universities Physics Alliance}, Institute for Astronomy, University of Edinburgh, Royal Observatory, Blackford Hill,
Edinburgh EH9 3HJ, UK}
\maketitle
\vspace {7cm}

\begin{abstract}
We present detailed clustering measurements for a flux limited 
sample of $\sim 14,000$ quasars extracted from the 2dF QSO Redshift Survey 
(2QZ) in the redshift range $0.8<z<2.1$.
After splitting the sample into three redshift bins and each of them
into six luminosity intervals, we estimate the quasar projected auto and 
cross-correlation functions at a given redshift for separations 
$3 \simlt r/{h^{-1} {\rm Mpc}}\simlt 20$.
Fitting the data with a biased CDM model and using a frequentist analysis 
(the $F$-test), 
we find that models with
luminosity dependent clustering are statistically favoured 
at the 95 per cent confidence level for $z>1.3$.
On the other hand, a number of tests based on information theory and
Bayesian statistics show only marginal evidence for luminosity
dependent clustering.
Anyway, the quality of the data is not good enough to accurately quantify 
how quasar biasing depends on luminosity.
We critically discuss the limitations of our dataset and show that a much
larger sample is needed to rule out current models for luminosity segregation.
Studying the evolution of the clustering amplitude with redshift,
we detect
an increase of the quasar correlation length with lookback time at the
99.3 per cent confidence level. 
Adopting the concordance cosmological model,
we discuss the evolution of quasar biasing with cosmic epoch and show that
quasars are typically hosted by dark matter haloes with mass 
$\sim 10^{13} M_\odot$.
\end{abstract}


\section{Introduction}
It is widely believed that
quasars are powered by accretion onto supermassive black holes.
However, 
a detailed understanding of the physical processes leading to quasar 
activity (and their connection with galaxy formation) is still lacking.

Simple semi-analytic models associate quasars with galaxy major mergers
and assume 
a tight relation between their instantaneous luminosity and the mass 
of the central black-hole, $M_{\rm bh}$
(Kauffmann \& Haehnelt 2000; Wyithe \& Loeb 2003; Volonteri et al. 2003).
The fraction of gas accreted onto the black hole during each merger 
is chosen to match the observed relation
between the velocity dispersion of the bulge and $M_{\rm bh}$
(Ferrarese \& Merritt 2000).
This ends up producing a correlation between 
the quasar luminosity and the mass of the host dark-matter halo.
Since the clustering properties
of dark-matter halos strongly depend on their mass, the quasar
clustering amplitude is thus expected to sensibly depend on luminosity.

Recent numerical simulations of galaxy mergers including black-hole accretion 
and feedback have cast some doubts on this picture (Springel et al. 2005).
These numerical experiments suggest that
a given black-hole produces quasar activity with a wide
range of luminosities (Hopkins et al. 2005). 
During its active phase, the black-hole is most likely 
observed as a relatively low-luminosity quasar with a small Eddington ratio.
For a short period of time, however, its emission reaches its peak value
(close to the Eddington luminosity)
which is indeed proportional to the mass of the powering black-hole.
Based on these models,
Lidz et al. (2006) conclude that quasar clustering should depend only
weakly on luminosity. 

From the observational point of view,
only recently quasar samples have grown big enough 
(in terms of number of objects)
to attempt the study of the clustering amplitude as a function of luminosity.
By analyzing the galaxy-quasar cross-correlation at $1.8\simlt z\simlt 3.5$
Adelberger \& Steidel (2005) found no evidence for luminosity-dependent 
clustering. They used 79 quasars spanning 4.4 orders of magnitude in absolute
luminosity which have been divided into 2 luminosity bins.
Larger samples are obviously required to confirm this result.

Croom et al. (2002, 2005) studied the redshift-space clustering amplitude of 
2dF quasars as a function of their apparent magnitude. 
Even though these authors initially found weak evidence for brighter QSOs 
being more 
strongly clustered, their most recent analysis shows no indication of
luminosity-dependent clustering.
These two studies, however, 
do not address the issue of luminosity dependent clustering.
In fact, they consider magnitude-limited samples within a broad 
redshift range ($0.3<z<2.9$ corresponding to $\sim 8$ Gyr in the currently
favoured cosmological model) 
and totally ignore any changes in the clustering signal as a
function of cosmic epoch. Moreover, their redshift-space analysis
complicates the interpretation of the clustering amplitudes, as 
the effect of non-linear peculiar velocities
could also depend on luminosity and redshift.
For these reasons, their null result 
does not necessarily imply that quasar clustering is 
independent of luminosity.

In this paper, which is a follow up analysis of Porciani, 
Magliocchetti \& Norberg (2004, PMN04 hereafter), 
we study the clustering properties of
$\sim 14,000$ quasars extracted from the complete 
2dF QSO Redshift Survey (Croom et al. 2004).
Our goal is to accurately
measure the real-space clustering amplitude of 2dF quasars 
as a function of redshift and absolute luminosity. 
To do this, we first split our quasar sample into three 
redshift bins and each subsample into six luminosity 
intervals. We then compute, in each redshift range, the 
associated projected auto- and cross-correlation functions, 
which, by construction, are not affected by redshift space 
distortions.

Using the largest quasar sample presently available, our 
results reveal a statistically significant evolution 
of the clustering length with redshift (as found already in PMN04)
but only weak evidence for luminosity dependent quasar clustering.

The layout of the paper is as follows.
In Section~\ref{sec:data} we describe our quasar 
sample and how we split it both in redshift and in luminosity.
In Section~\ref{sec:xiproj}, we measure
the projected auto- and cross-correlation functions of the different
subsamples. 
The issue of luminosity dependent clustering is addressed in 
Section~\ref{sec:bias}. 
Using a Monte Carlo Markov Chain, we
estimate the quasar clustering amplitude as a function of 
redshift and luminosity.
A number of robust tests are then used to evaluate the statistical significance
of the measured luminosity dependence in a given redshift range.
A critical discussion of the limitations of our data 
and of possible future strategies is also presented here.
In Section~\ref{sec:redshift}, we focus on the pure redshift dependence of
the quasar clustering amplitude and we provide fitting functions
for the evolution of the quasar correlation length and bias with cosmic epoch.
All our results are summarized in Section~\ref{sec:discussion}.

Throughout this paper we assume a ``concordance'' cosmological model with
mass density parameter $\Omega_0=0.3$
(with a baryonic contribution $\Omega_b=0.049$), vacuum energy density
parameter $\Omega_\Lambda=0.7$ and present-day value of the Hubble constant
$H_0=100\, h\,{\rm km}\,{\rm s}^{-1}\,{\rm Mpc}$ with $h=0.7$. 
We also adopt a cold dark matter (CDM) power
spectrum with primordial spectral index $n=1$ and with normalization 
fixed by $\sigma_8$, the rms linear density fluctuation within a 
sphere with radius of $8$~\Mpc.

\section{Quasar sample definition}
\label{sec:data}

The 2dF QSO Redshift Survey (2QZ) includes 23,338 quasars 
which span a wide redshift range ($0.3\simlt z \simlt 2.9$)
and are spread over 721.6 deg$^{2}$ on the sky
(see Croom et al. 2004).
In order to minimise systematic effects, we restrict our analysis 
to regions with spectroscopic (photometric) completeness larger 
than 70 (90) per cent, which limits the redshift range to 
$0.5<z<2.1$. Only quasars brighter than 
M$_{b_{\rm J}}-5\log_{10}h=-21.7$ are considered, which ensures 
the exclusion of quasars for which the contribution from the
host galaxy may have led to a mis-identification of the source.

In order to make a physically motivated analysis and isolate 
evolutionary effects, we subdivide our sample 
into three redshift bins. As in PMN04, we require a similar number of quasars 
to lie in each redshift bin and that each sub-sample covers 
a not too different interval of cosmic time.
To better satisfy these criteria (see PMN04 fur further details),
we impose an additional redshift cut so as to keep 
only quasars within $0.8<z<2.1$. With this 
selection, we end up with nearly 14,000 2QZ quasars that we split into
the redshift intervals: $0.8<z<1.3$, $1.3<z<1.7$ and $1.7<z<2.1$ 
(containing each between $\sim~4300$ and $\sim~4900$ quasars).
%

To study the dependence of the quasar clustering amplitude on luminosity,
we further divide each sub-sample into six 
complementary sets based on quasar absolute luminosity: (F25, B75), 
(F50, B50) and (F75, B25), where F$x$ and B$y$ correspond to 
the $x$ per cent faintest quasars and the $y$ per cent brightest quasars
(in terms of their absolute magnitude).
Table~\ref{tab:data} 
lists the main properties of each sample, including the number 
of quasars (col. 4), the median redshift (col. 5), the absolute 
magnitude range (cols. 6 \& 8) and the sample median absolute 
magnitude (col. 7).

\begin{table}
\begin{center}
\caption{Main properties of our data sets. The subscripts min, max and med
respectively denote the minimum, maximum and  
median value of a variable.
\label{tab:data}}
\begin{tabular}{cccccccc}
\hline
\hline
$z_{\rm min}\!\!\!\!$&$z_{\rm max}\!\!\!\!$& Data & N$_{\rm QSO}$ &$z_{\rm med}\!\!\!\!$ &
$M_{\rm max}\!\!\!\!$& $M_{\rm med}\!\!\!\!$ & $M_{\rm min}\!\!\!\!$ \\
& & & & & \multicolumn{3}{c}{$M_{b_{\rm J}}-5\log_{10}h$} \\
\hline
0.8 & 1.3 & F25 & 1232 & 0.93& -21.7 & -22.4 & -22.7 \\
0.8 & 1.3 & F50 & 2464 & 0.99& -21.7 & -22.7 & -23.2 \\
0.8 & 1.3 & F75 & 3696 & 1.04& -21.7 & -22.9 & -23.7 \\
0.8 & 1.3 & B75 & 3698 & 1.12& -22.7 & -23.4 & -25.3 \\
0.8 & 1.3 & B50 & 2466 & 1.14& -23.1 & -23.6 & -25.3 \\
0.8 & 1.3 & B25 & 1234 & 1.15& -23.6 & -24.0 & -25.3 \\
\hline
1.3 & 1.7 & F25 & 1285 & 1.44& -22.8 & -23.3 & -23.5 \\
1.3 & 1.7 & F50 & 2571 & 1.48& -22.8 & -23.5 & -23.9 \\
1.3 & 1.7 & F75 & 3857 & 1.49& -22.8 & -23.6 & -24.3 \\
1.3 & 1.7 & B75 & 3858 & 1.53& -23.4 & -24.0 & -26.0 \\
1.3 & 1.7 & B50 & 2572 & 1.53& -23.8 & -24.3 & -26.0 \\
1.3 & 1.7 & B25 & 1286 & 1.55& -24.3 & -24.7 & -26.0 \\
\hline
1.7 & 2.1 & F25 & 1081 & 1.83& -23.4 & -23.8 & -24.0 \\
1.7 & 2.1 & F50 & 2163 & 1.86& -23.4 & -24.0 & -24.3 \\
1.7 & 2.1 & F75 & 3244 & 1.87& -23.4 & -24.1 & -24.9 \\
1.7 & 2.1 & B75 & 3245 & 1.91& -23.9 & -24.5 & -26.4 \\
1.7 & 2.1 & B50 & 2163 & 1.91& -24.3 & -24.8 & -26.4 \\
1.7 & 2.1 & B25 & 1082 & 1.92& -24.8 & -25.2 & -26.4 \\
\hline
\hline
\end{tabular}
\end{center}
\end{table}

\begin{figure*}
\centerline{\epsfig{figure=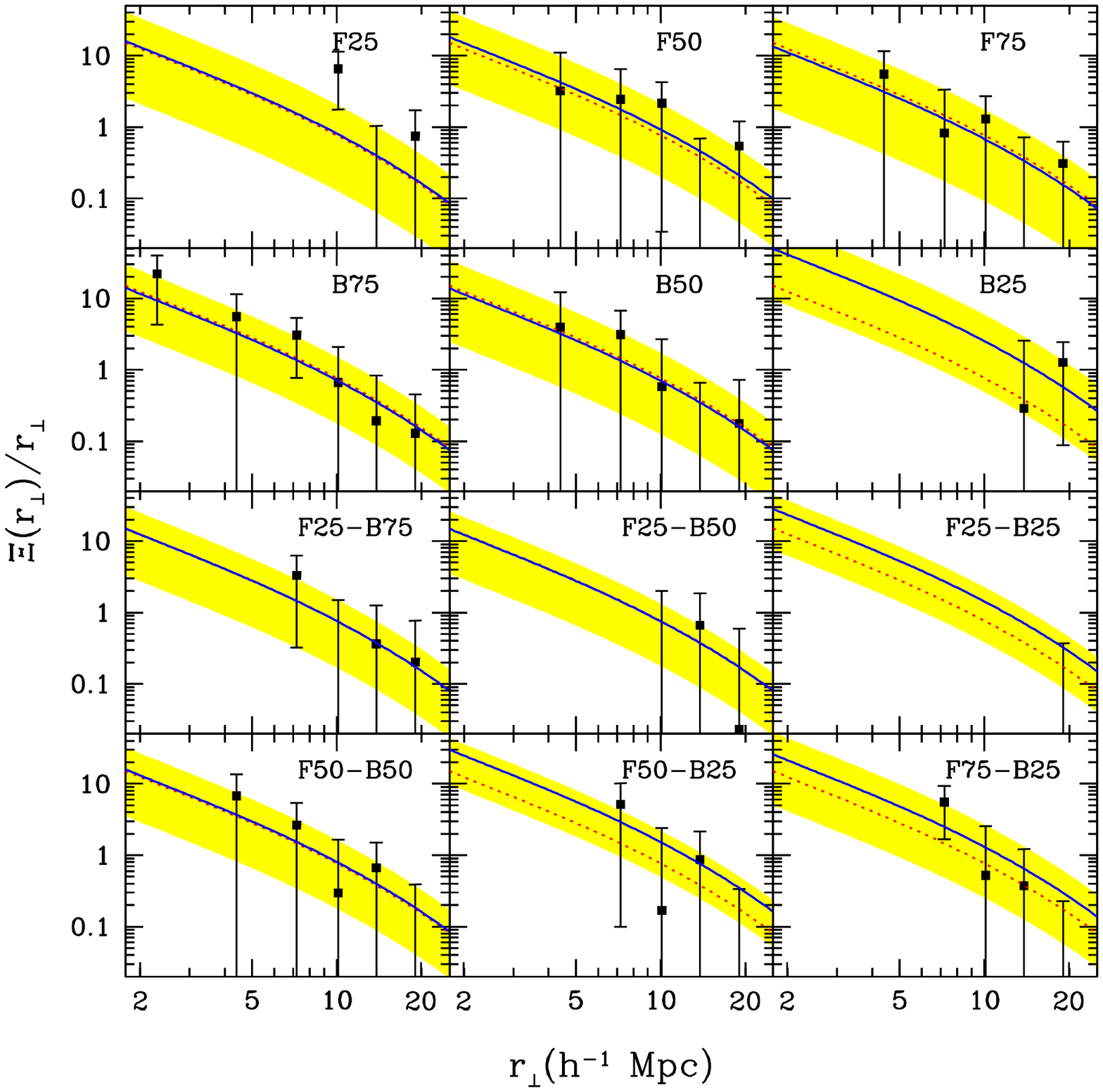,height=8cm}
\hspace*{-0.3cm}
\epsfig{figure=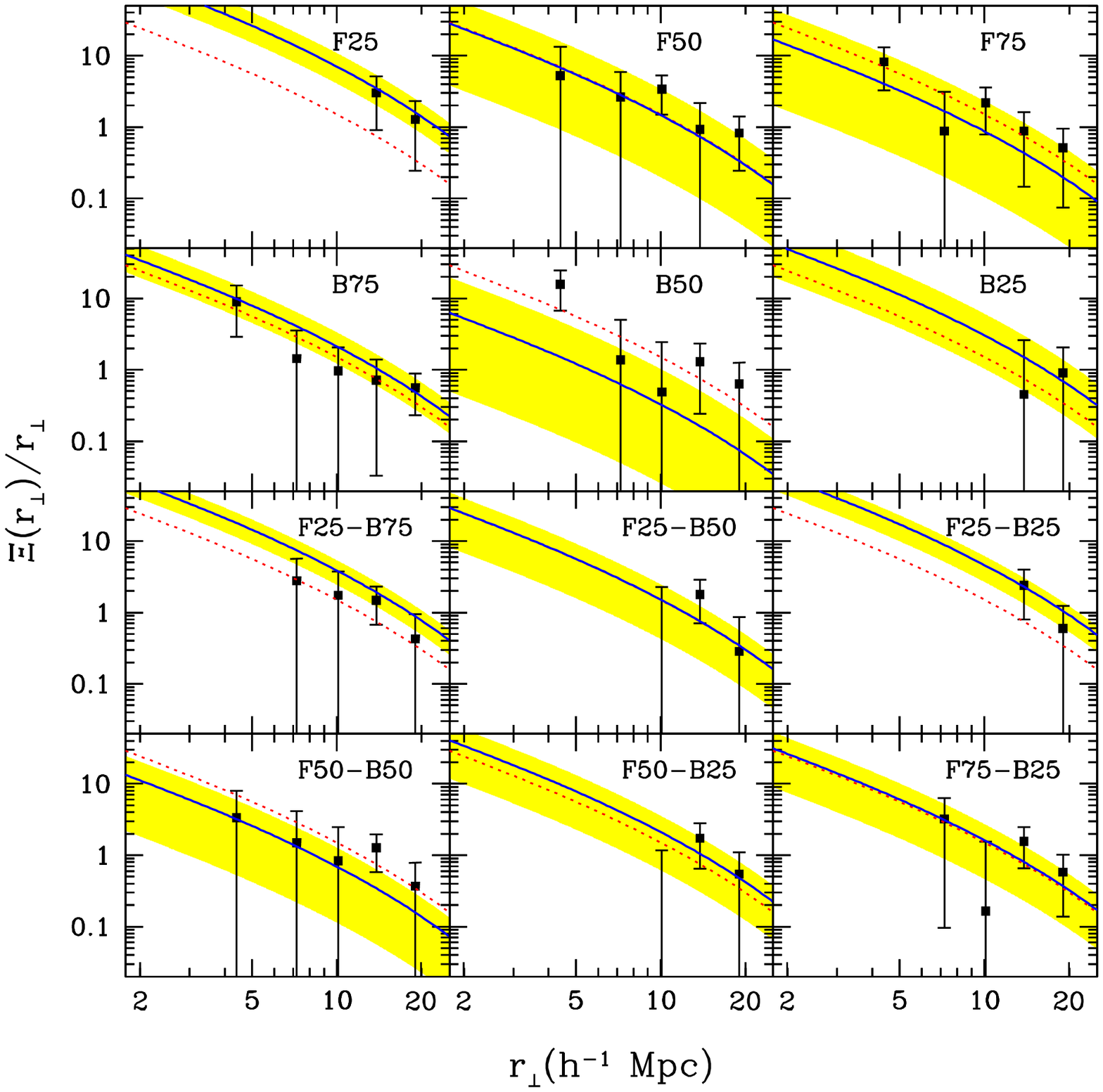,height=8cm}}
\centerline{
\epsfig{figure=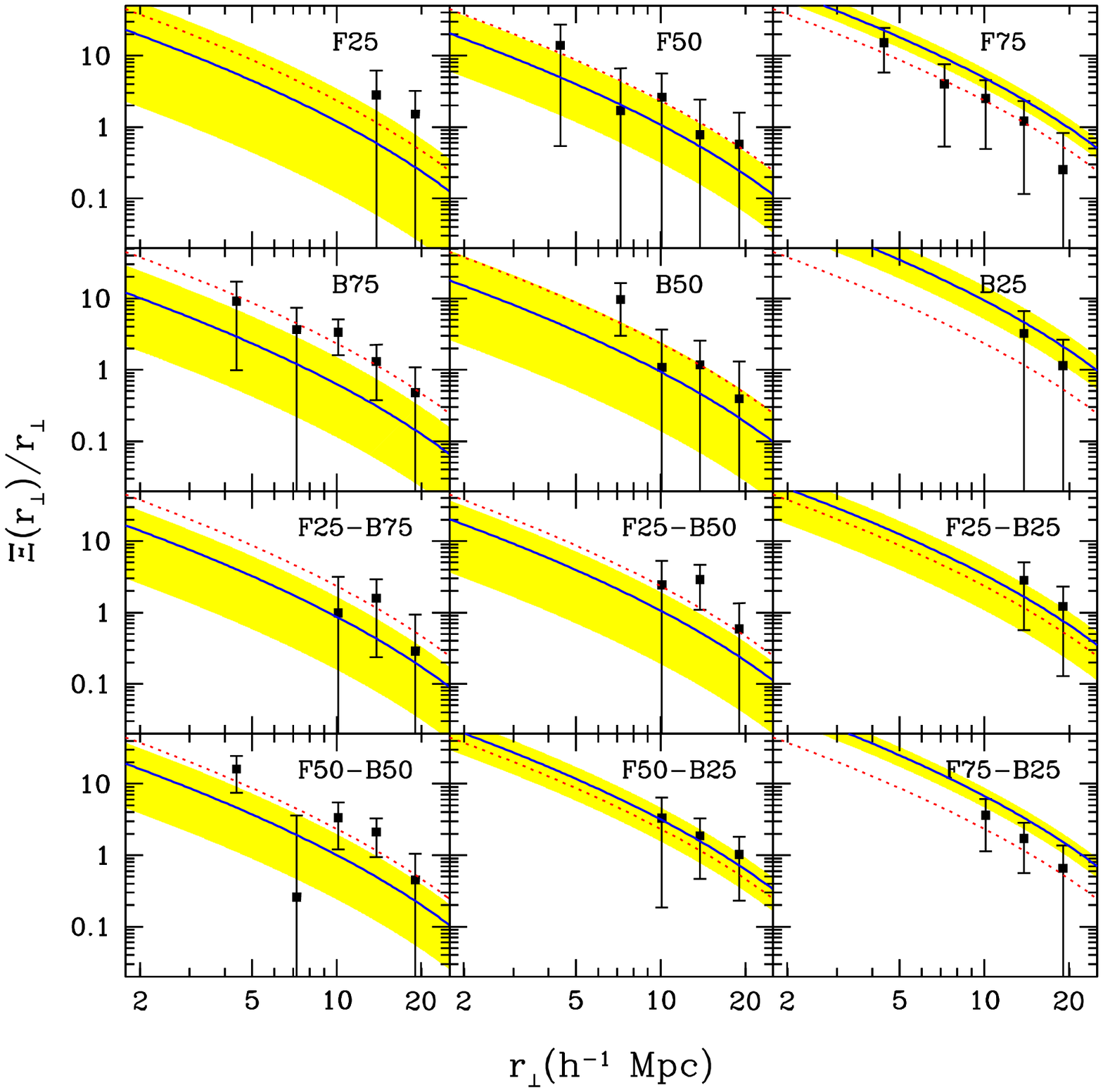,height=8cm}\hspace*{-0.3cm}
\epsfig{figure=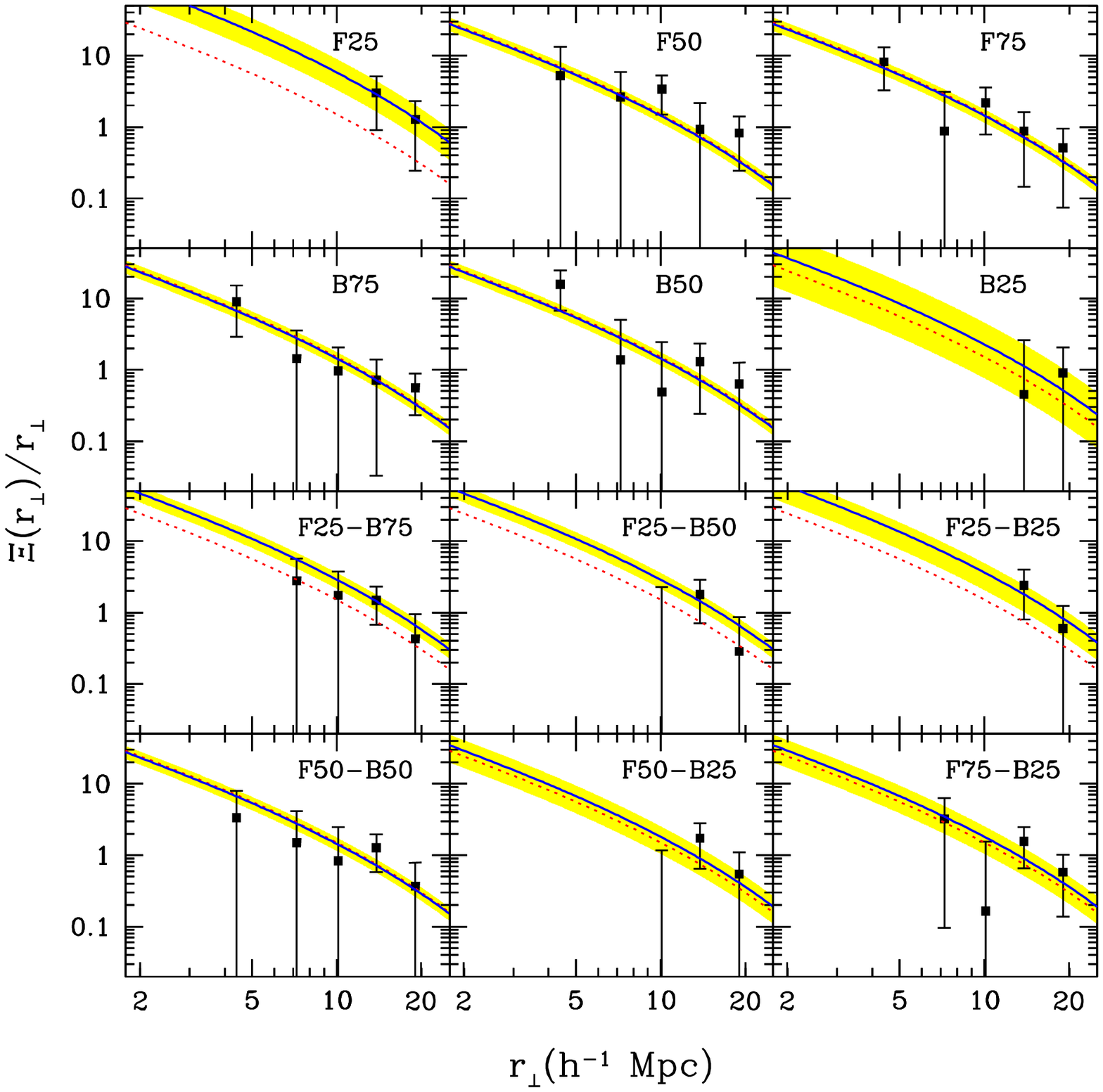,height=8cm}}
\caption{{\it Top-left panel:}
The projected auto-correlation function (top two rows) and
cross-correlation function (bottom two rows) for the low-redshift 
quasar samples ($0.8<z<1.3$). Only measurements obtained with more 
than 40 quasar pairs are plotted (filled squares with errorbars).
Bins with less pairs are omitted and not considered for model fitting. 
For each luminosity sub-sample, the shaded band show the
central 68 per cent range of 
best-fitting correlation functions (assuming that each luminosity interval
is associated with a different bias parameter). 
The mean value of the marginalized bias distribution
is represented with a solid line. For reference, we also plot 
the best-fitting correlation function for the 
entire low-redshift sample derived in PNM04 (dotted line).
{\it Top-right and bottom-left panels:} Same as the top-left panel, but for 
the medium- ($1.3<z<1.7$) and high- ($1.7<z<2.1$) redshift quasar 
samples respectively. 
{\it Bottom-right panel:} As in the top-right panel but using a three-parameter
model where only the B25 and F25 sub-samples have  different bias parameters
from all the others.
\label{fig:data}
}
\end{figure*}

\section{Quasars Clustering Analysis}
\label{sec:xiproj}

\subsection{Estimating the correlation functions}

The simplest statistic which can be used to quantify
clustering in the observed quasar distribution
is the 2-point correlation function in redshift space, 
$\xi^{\rm q}(r_\perp,\pi)$.
To measure this quantity we first
build a catalogue of
unclustered points which has the same angular and
radial selection function as the data. 
The radial selection function is obtained by
heavily smoothing the observed quasar comoving 
distance distribution, ${\cal N}(r)$. 
The quasar auto-correlation function is then 
estimated by comparing the probability distribution 
of quasar and random pairs on a two-dimensional 
grid of separations $(r_\perp, \pi)$.
We use both the Landy-Szalay estimator (Landy \& Szalay 1993) 
and the Hamilton estimator (Hamilton 1993):
\begin{equation}
\xi_{\rm{LS}}^{\rm q}= \frac{ DD - 2DR + RR }{RR}\;,\ \ \ 
\xi_{\rm{H}}^{\rm q} = \frac{DD\cdot RR}{(DR)^2}\,-\,1
\end{equation}
where $DD$, $DR$ and $RR$ are the suitably normalised numbers of  
weighted data-data, data-random and random-random pairs in 
each bin.\footnote{Note that, in this case, there is no 
need to use the standard $J_3$ (minimum variance) weighting 
scheme since the mean density of quasars,
$n_{\rm q}$, is so low that $1\,+\,4\,\pi\,J_3\,
n_{\rm q}\simeq 1$ for any reasonable quasar clustering
amplitude.}
As expected, the two estimators give comparable 
answers within the errors. In what follows 
we only present results obtained with the Landy-Szalay estimator.
Similarly, we use the appropriately symmetrized Landy-Szalay 
estimator to compute the quasar cross-correlation function:
\begin{equation}
\xi_{\rm{LS}}^{\rm q}= \frac{ D_iD_j - (D_iR_j+D_jR_i) + R_iR_j }{R_iR_j}\;,\ \ \ 
\end{equation}
where 
$D_iD_j$, $D_iR_j$, $D_jR_i$ and $R_iR_j$ are the suitably normalised 
numbers of weighted data and random pairs from samples $i$ and $j$.

To avoid redshift space distortions, 
and determine the quasar clustering amplitude 
in real space, one can then use the `projected correlation 
function' which is obtained by integrating 
$\xi^{\rm q}(r_\perp,\pi)$ in the $\pi$ direction:
\begin{equation} 
\frac{\Xi^{\rm q}(r_\perp)}{r_\perp} =  
\frac{2}{r_\perp} \int_{0}^{\infty} 
\xi^{\rm q}(r_\perp,\pi)\,{\rm d}\pi .
\label{proj}
\end{equation}
With the current quasar sample, we find that a reliable measure of
$\xi^{\rm q}(r_\perp,\pi)$ is only achievable on scales
$\pi \simlt 50$~\Mpc. In order to avoid the measured signal 
to be dominated by noise, we
limit the integration to an upper limit, $\pi_{\rm max}$.
As explained in PMN04, $\pi_{\rm max}=45$~\Mpc\ is an adequate 
choice, and we have checked that our results are not sensitive 
to the precise value adopted.

As in PMN04, we use a blockwise bootstrap resampling technique 
to estimate the uncertainties of our clustering measurements. 
For each redshift interval, we divide each of the two 2QZ areas
into 28 equal-volume regions, containing, within a 
factor of two, the same number of quasars (the minimum size
of a subvolume is nearly $180 h^{-1}$ Mpc for the low-redshift
sample, $250 h^{-1}$ Mpc for the medium-redshift
sample and $220 h^{-1}$ Mpc for the high-redshift;
since we measure correlations on scales smaller than $20 h^{-1}$ Mpc
our error estimates should hardly be affected by edge-effects and/or missing
large-scale structure). \footnote{We have also checked that doubling the
size of the subvolumes does not significantly alter the diagonal errors.}
We build hundreds of bootstrap-samples (see \S \ref{fit}), 
each of them composed by 56
sub-samples (28 for each 2QZ region) randomly drawn (allowing 
repetitions) from the set described above. We measure the 
projected auto- and cross-correlation functions for all the
bootstrap samples. 
For each 
$r_\perp$, we identify the {\it rms} variation of $\Xi^{\rm q}$ over 
the bootstrap-samples with the 1$\sigma$ error for the 
projected auto- and cross-correlation function.
Note the our bootstrap uncertainties account for both sampling
and estimation errors.

Our method for estimating errors relies on the fact that 
our dataset is statistically representative of the quasar 
distribution in the Universe. 
\footnote{Our blockwise bootstrap errors are in
good agreement with those obtained using
mock galaxy catalogues extracted from high-resolution numerical simulations.
Discrepancies between the two types of errors are of order 20 per cent or
less on the scales used in this paper.}
However, this cannot be true 
for bins of spatial separations which contain just a few 
quasar pairs. Therefore, in what follows, we ignore 
clustering results obtained with less than 40 quasar 
pairs. Depending on the sample, this corresponds to 
$r_\perp<2-6$~\Mpc.

Our results are presented in Fig.~\ref{fig:data} 
(filled squares with errorbars).
All the correlations in a given redshift bin have similar amplitudes and
are rapidly decreasing with $r_\perp$.
As expected, errorbars are smaller for the largest quasar samples.

\subsection{Modelling the data}
\label{sec:data_mod}

Given the relatively large errorbars, it is hard to spot any luminosity
dependence of quasar clustering by simply looking at 
the correlations in Fig.~\ref{fig:data}.
For this reason, we use a reference model to fit the data.
This is obtained by assuming that each sample is characterized by 
a linear bias parameter which does not depend on spatial separation within
the range of scales we analyse.
In other words, we assume that
the quasar 2-point auto- and cross-correlations scale as
\begin{equation}
\xi^{\rm q}_{ij}(r) = b_{i}\,b_{j}\, \xi(r) \;,
\label{eq:fit}
\end{equation}
where $i$ and $j$ are the labels of 2 quasar sub-samples 
within the same redshift 
bin,
$r$ denotes the comoving separation between quasar pairs, 
and $b_i$ is the bias parameter of the $i$-th sub-sample 
with respect to the mass 
autocorrelation function, $\xi(r)$, computed as in Peacock \& Dodds (1996, 
nearly undistinguishable results 
are found using the method by Smith et al. 2003) 
assuming a value for $\sigma_8$.
\footnote{Note that samples with different luminosities have 
slightly different median redshifts. However the variation of 
the mass correlation function among them is of the order of a 
few per cent, much smaller than the uncertainty in the quasar 
clustering amplitude. This implies
that we can safely use the same $\xi(r)$ (evaluated at the median redshift of 
the redshift bin) for all the luminosity subsamples. We have tested that
this does not influence our results.}
Within the framework of halo models 
(see PMN04 for a direct application to quasars), the 
assumption that the amplitude  
of the cross-correlation function between haloes of different 
masses scales as the geometric mean of the correlations 
of the individual haloes is very natural in the two-halo regime. 
We have checked against numerical
simulations that, for dark matter haloes, 
the assumption holds over the redshift range 
and scales considered here, i.e. $3\simlt r \simlt\,20$~\Mpc. 
The corresponding projected correlation functions 
are obtained through this simple integral relation:
\begin{eqnarray}
\Xi^{\rm q}_{ij}(r_\perp) &=& 2\,b_{i}\,b_{j} \int_{r_\perp}^{\infty}\frac{r\,\xi(r) }{(r^2-r_\perp^2)
^{1/2}}\,{\rm d}r\;.
\label{eq:xibar}
\end{eqnarray}

\subsection{Fitting the data with correlated errors}
\label{fit}
We use a minimum least-squares method 
(corresponding to a maximum likelihood method in the case 
of Gaussian errors) to determine the bias parameters that 
best describe the clustering data. 
In each redshift range, we sample
the six-dimensional parameter space (one bias parameter per luminosity
range) using a Markov Chain Monte Carlo method.
A principal 
component analysis (see e.g. Porciani \& Giavalisco 2002) 
is used here to deal with correlated errorbars. 
The principal components of the errors are computed 
by diagonalizing the covariance matrix obtained by resampling 
the data with the bootstrap method. The objective function 
(the usual $\chi^2$ statistic) is then obtained by only considering 
the most significant principal components (i.e. those contributing
the largest fraction of the total variance). 
\footnote{The covariance matrix obtained from bootstrap resampling
is only an estimate of the true one and contains an intrinsic uncertainty.
The errors in its component propagate in the calculation of its eigenvalues
and eigenvectors.
It is therefore recommended to consider only the principal components 
corresponding to the largest eigenvectors in the fitting procedure
(see \S4.2 in Porciani \& Giavalisco (2002) for further details).}
Unfortunately there is no unique objective way of deciding 
how many principal components should be considered for a given dataset.
Considering too few components (and thus discarding information contained
in the data), one obtains too good values for the objective function 
(i.e. $\chi^2_{\rm min}/{\rm dof} \ll 1$, where dof indicates the number
of degrees of freedom)
that correspond to unrealistically large uncertainties for the 
fitted parameters. On the other hand, considering too many 
components (and thus, most likely, introducing noise),
one often obtains bad fits (i.e. $\chi^2_{\rm min}/{\rm dof} \gg 1$) 
that correspond to unrealistically small errors for the bias parameters.
In factor analysis,
a number of empirical methods (e.g. Kaiser criterion, scree test) are often
employed to select the number of significant components.
The robustness of these techniques for model fitting is, however, rather weak.
We thus decided to select the number of components to account
for a fixed fraction (95 per cent) of the bootstrap variance.
We have checked that this compression 
guarantees a detailed reconstruction of the model-data residuals
and simultaneously avoid the $\chi^2$ to be dominated 
by deviations along the components corresponding to tiny eigenvalues.
Moreover, the reduced $\chi^2$ of the best-fitting models obtained this way 
is of order unity in all cases, as expected theoretically. 
In what follows,
we thus use the symbol $\chi^2_{95}$ to denote
the $\chi^2$ function computed by considering
the first principal components that, in total, account for $95$ per cent of the
variance. This corresponds to 19 principal components (out of 42 datapoints)
for the high-redshift sample, 21 (out of 45) for the median-redshift sample
and 22 (out of 47) for the low-redshift sample.

Since the bootstrap technique is very time consuming, a careful choice
of the number of resamplings is required.
For this reason,
we performed a number of Monte Carlo simulations checking for the
stability of eigenvalues, eigenvectors and $\chi^2_{95}$ estimates.
In practice, we first bootstrapped our data and computed the corresponding
covariance matrix $C_{ij}$. We then built a large number of realizations
of a Gaussian process with covariance $C_{ij}$. 
Finally, we estimated the covariance matrix of the Gaussian process
from a finite number of realizations, $N$, and studied its dependence on $N$.
We found that a few hundred bootstrap resamplings are needed for a robust
estimation of the $\chi^2_{95}$ function.
To overcome the Gaussian hypothesis,
we also studied the convergence of the covariance matrix obtained
by directly bootstrapping our data.
We found that
estimates of $\chi^2_{95}$ converge (i.e. 
present negligible scatter) when nearly 250 bootstrap resamplings are used.
This has been obtained by using 500 resamplings of the median redshift sample. 
To be on the safe side, 
we thus decided to use at least 350 bootstrap resamplings for each redshift 
bin. We note that this implies over 12,000 correlation
function estimates, with each containing at least 50,000 random points.

As an additional test of the robustness of our results, we checked, 
a posteriori, how much our best-fitting models  
depend on the number of adopted components, $N_{\rm pca}$. 
This is discussed in detail in Section \ref{sec:bias} 
where we present our results.

\begin{figure*}
\centerline{\epsfig{figure=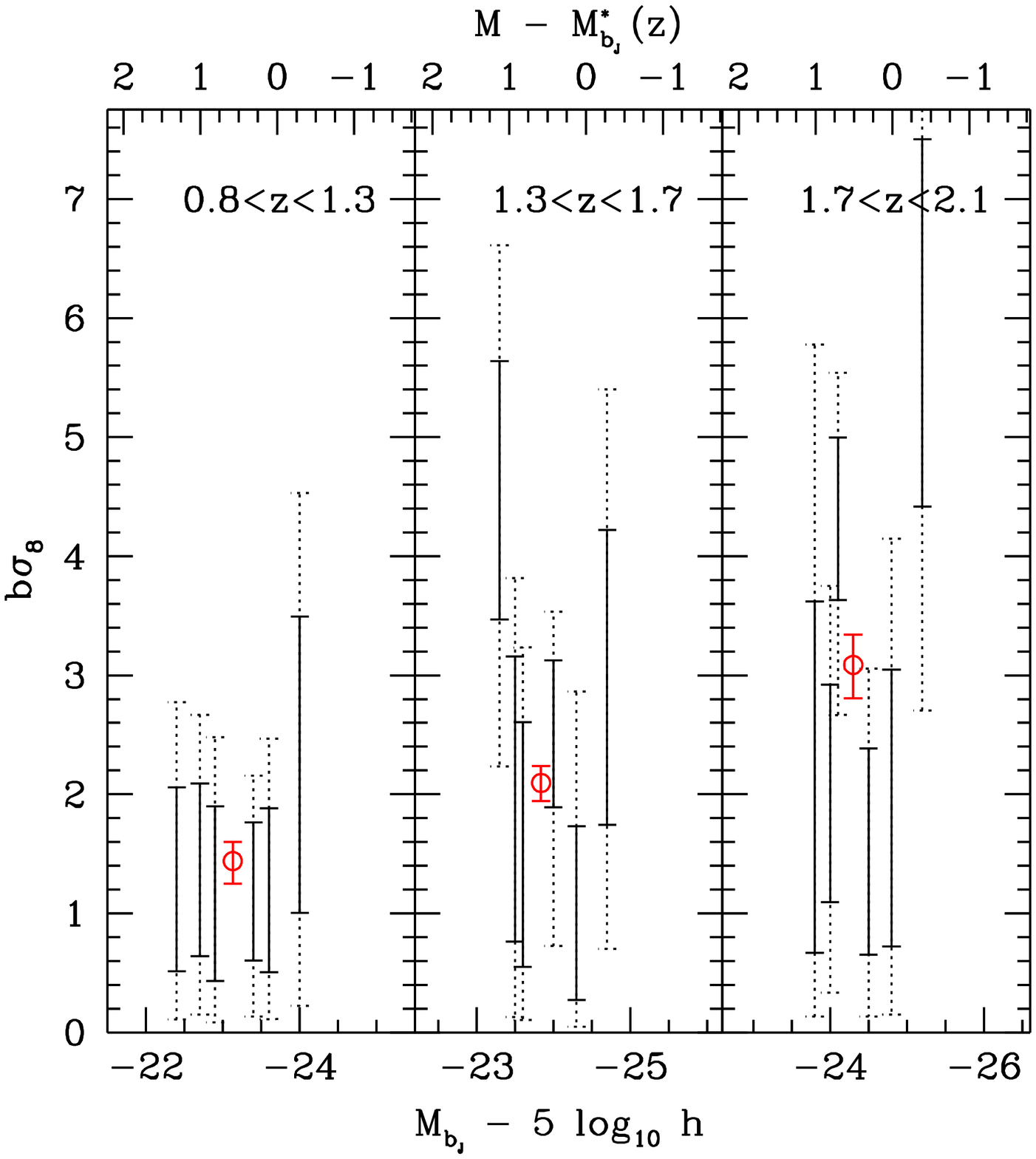,width=0.48\textwidth}
\epsfig{figure=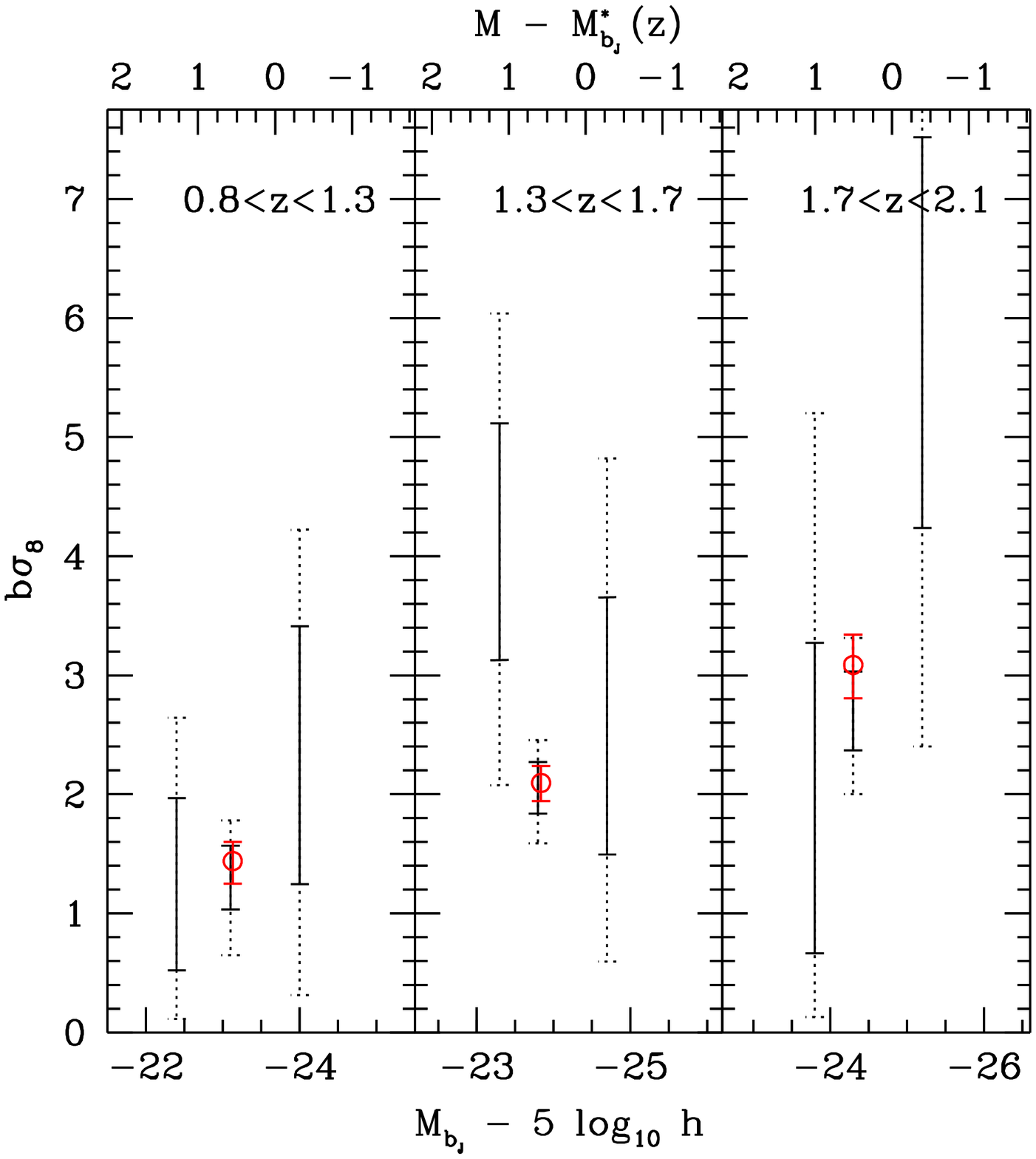,width=0.48\textwidth}}
\caption{{\it Left panel:} $\sigma_8^{\rm q}$, i.e. $b \sigma_8$, as 
function of the median absolute magnitude 
for the low (left), medium (centre) and high (right) redshift bins. For each 
sample, the errorbars indicate the central 68.3 (solid) 
and 95.4 (dotted) credibility intervals of 
the marginalized bias probability distribution. 
On the upper axis, we 
indicate the magnitude difference with respect to \mstar(z), where 
the later is given by Eq.~13 of Croom \etal\ (2004). The open 
symbol show the result obtained by fitting the correlation function of
all the quasars lying in this redshift bin (see PMN04).
{\it Right panel:} Same as the left panel but for a three-parameter fit (see
text).
\label{fig:bias_l}}
\end{figure*}

\section{Luminosity dependent clustering}
\label{sec:bias}

In this section, we present the results obtained by fitting
the auto- and cross-correlation functions presented
in Fig.~\ref{fig:data} with the model given in equation (\ref{eq:fit}).
Our aim is to quantify how the bias parameter of optically bright
quasars depends on redshift and luminosity.

The best-fitting functions obtained with 
the Markov Chain Monte Carlo method are overplotted to the data 
in Fig.~\ref{fig:data} (shaded regions). 
Note that, for all redshift 
bins, they accurately describe the quasar auto- and cross-correlations
in the whole range of separations under analysis.
In the left panel of Fig.~\ref{fig:bias_l}, 
we plot $\sigma_8^{\rm q}$ (i.e. $b \sigma_8$) as a function 
of absolute magnitude. 
As expected, the bias parameter of all quasars lying in a given redshift bin
(measured by PMN04 and represented with an open symbol) 
lies in between the values found here.
Taken at face value,
the data show some evidence for luminosity segregation; 
however, samples at
different redshifts show different trends.
The brightest quasars in 
the high-redshift bin seem to be more strongly clustered than the others.
In the medium-redshift bin,  the bias seems to follow a $U$-shape:
the faintest and the brightest quasars in the set are typically more 
strongly clustered than \mstar\ quasars. 
On the other hand, the low-redshift bin does not show any particular trend:
the bias keeps constant with luminosity.

Note, however, that correlations between errorbars in Fig.~\ref{fig:bias_l} 
might create spurious trends with luminosity.
Moreover, all the different sub-samples in a given redshift bin 
have consistent bias parameters at the $2\sigma$ level.
Given these concerns, we want to use a robust statistical test to
investigate whether our data show any sign of luminosity dependent
clustering.

\subsection{Does quasar clustering depend on luminosity?}
\label{sec:stat}

If there is no luminosity segregation, all our data should
be described by one single bias parameter (i.e. all the $b_i$ 
should assume the same identical value). If, instead, some 
luminosity bins show a statistically significant deviation 
from the overall clustering amplitude, then a description with a 
number of different bias parameters should be preferred.
Therefore, in this section we address the issue of 
luminosity dependent clustering by answering the following 
question: how many bias parameters are required to adequately 
describe the quasar auto- and cross-correlations in a given 
redshift bin? 
In other words, we want to understand how many different bias parameters
can be reliably measured in a statistically significant way.
This is a classical problem of model selection where we want to
find the proper tradeoff between goodness of fit (in the $\chi^2$ sense) 
and complexity (in terms of number of free parameters).
We use five different methods: the F-test, the Akaike information
criterion (AIC), the Bayes factors (BF), the Bayesian information criterion 
(BIC) and
the deviance information criterion (DIC, see the Appendix for a brief review).
We consider three different models. In the first one, there is no
clustering segregation with luminosity and
all the luminosity bins at a given redshift are associated with the same 
bias parameter.
In the second model, we use
three bias parameters (one each for the brightest, B25, and 
faintest samples, F25, and one for the remaining quasars: hereafter 
those samples are referred to as B, M and F, for bright, medium 
and faint). 
In the third model, each luminosity bin has its own bias parameter,
as already described in the previous sections.
In Fig.~\ref{fig:bias_l}, we show that the results of the three- and 
six-parameters models are qualitatively and quantitatively similar.

Before comparing the different models, it is worth noticing that,
based on the $\chi^2$ statistic, a single 
parameter fit gives an acceptable description of the data for all redshift 
bins. Of course, using additional parameters reduces the 
minimum $\chi^2$ of the best-fitting models. 
We want now to understand whether this $\chi^2_{\rm min}$
reduction is statistically significant.

Performing the $F$-test and using the 95 per cent confidence
level as a threshold to prefer a model with respect to another,
we find that
the low-redshift sample is best-fitted by a one-parameter model, while
the medium- and high-redshift samples are best described 
by a six-parameter model.
Similarly, all the Bayesian and information-theory based tests mentioned above 
clearly indicate that the low-redshift 
sample is best described by a one-parameter model.
On the other hand, for the medium- and high-redshift samples,
the situation is more confused. Because of the different penalties
for model-complexity, depending on the adopted test,
either the six- or the one-parameter model is the preferred one.
However, with the exception of the six-parameter fit with the Akaike criterion,
no model can be rejected with high-confidence (see Table \ref{tab:model_sel} 
for a summary of the results).
Therefore, we conclude that, while there is no evidence of clustering
segregation with luminosity in the low-redshift sample,
we find marginal evidence for it
in the medium- and high-redshift samples.

\begin{table}
\begin{center}
\caption{Number of parameters of the preferred models
according to various model-selection criteria. 
When the evidence for the best model is not strong,
we list all the acceptable models (in order of decreasing evidence).
We use a 95 per cent confidence level for the F-test, AIC and DIC
while we discard models with substantial evidence according to
the Kass-Raftery criterion (see Appendix) 
for BF and BIC.
\label{tab:model_sel}}
\begin{tabular}{cccc}
\hline
\hline
Criterion & $0.8<z<1.3$& $1.3<z<1.7$& $1.7<z<2.1$ \\
\hline
F-test & 1 & 6 & 6 \\
AIC & 1-3 & 1-3   & 1-3\\
BF  & 1-3-6 & 6-3-1 & 6-3-1 \\
BIC & 1 & 1-3-6 & 1-3-6 \\
DIC & 1-3-6 & 6-3-1 & 6-3-1\\
\hline
\hline
\end{tabular}
\end{center}
\end{table}

Following a suggestion of the referee, 
we have also tried to combine
data from the different redshift bins by  
using a simple parameterization
of the luminosity and redshift dependent bias. 
For simplicity, we adopted a linear relation with quasar luminosity 
$b(M_{b_{\rm J}},z)=b_0(z)+b_1 10^{-0.4 (M_{b_{\rm J}}-M^{\star}_{b_{\rm J}}(z))}$.
In this case, the $68.3$ per cent confidence interval for
the luminosity-dependent parameter is $b_1=0.3^{+0.6}_{-0.7}$.
This confirms that the evidence for clustering segregation with luminosity
is marginal in our sample.
Similar conclusions are drawn adopting 
quadratic or cubic relations for the luminosity dependence of the bias
parameter. In all cases, the best-fitting model favours a larger bias parameter
for the brightest quasars in our sample. However, the statistical significance
of the result is low.

\subsection{Limitations of the PCA method}
\begin{figure*}
\centerline{\epsfig{figure=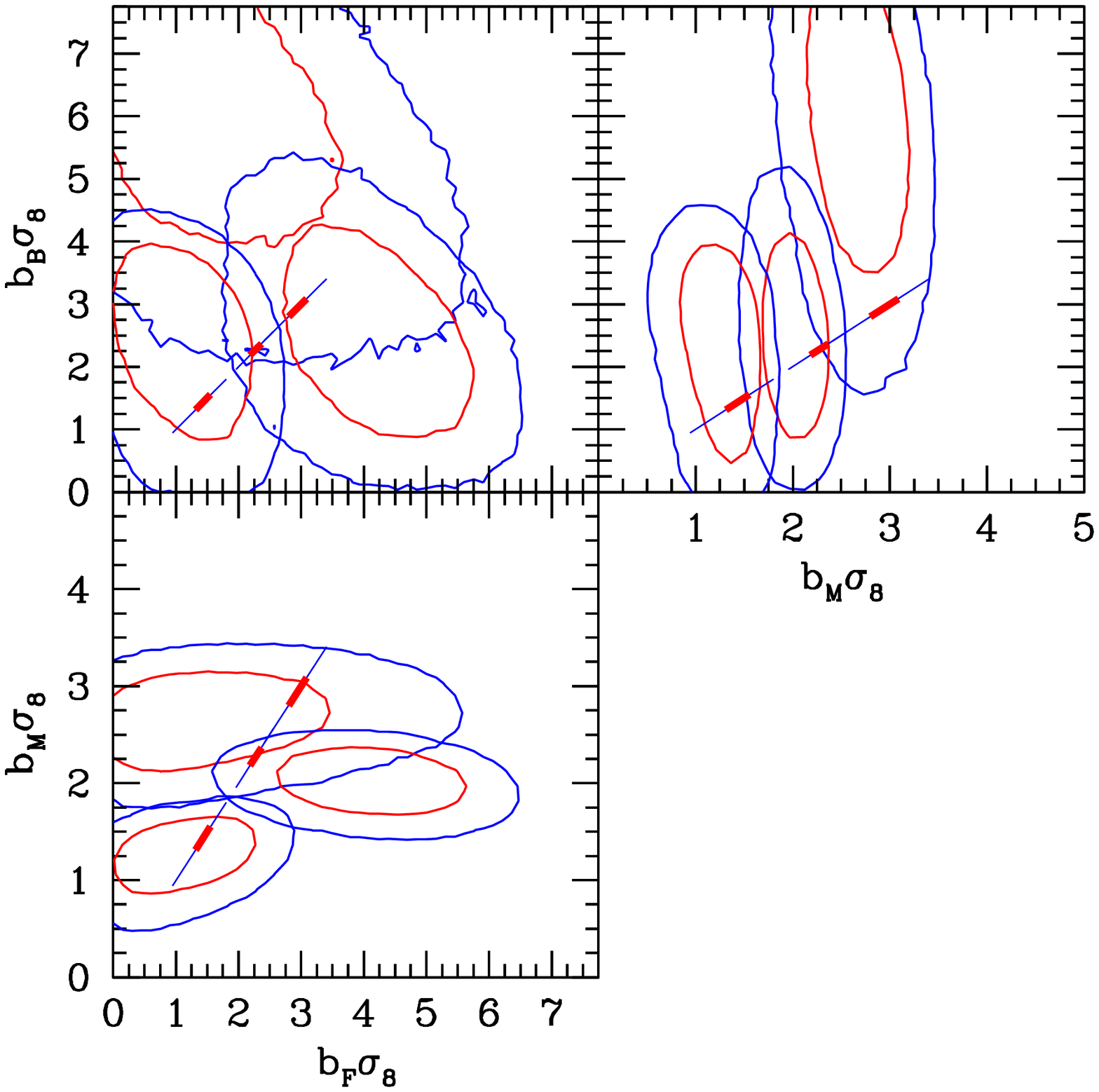,width=0.48\textwidth}
\epsfig{figure=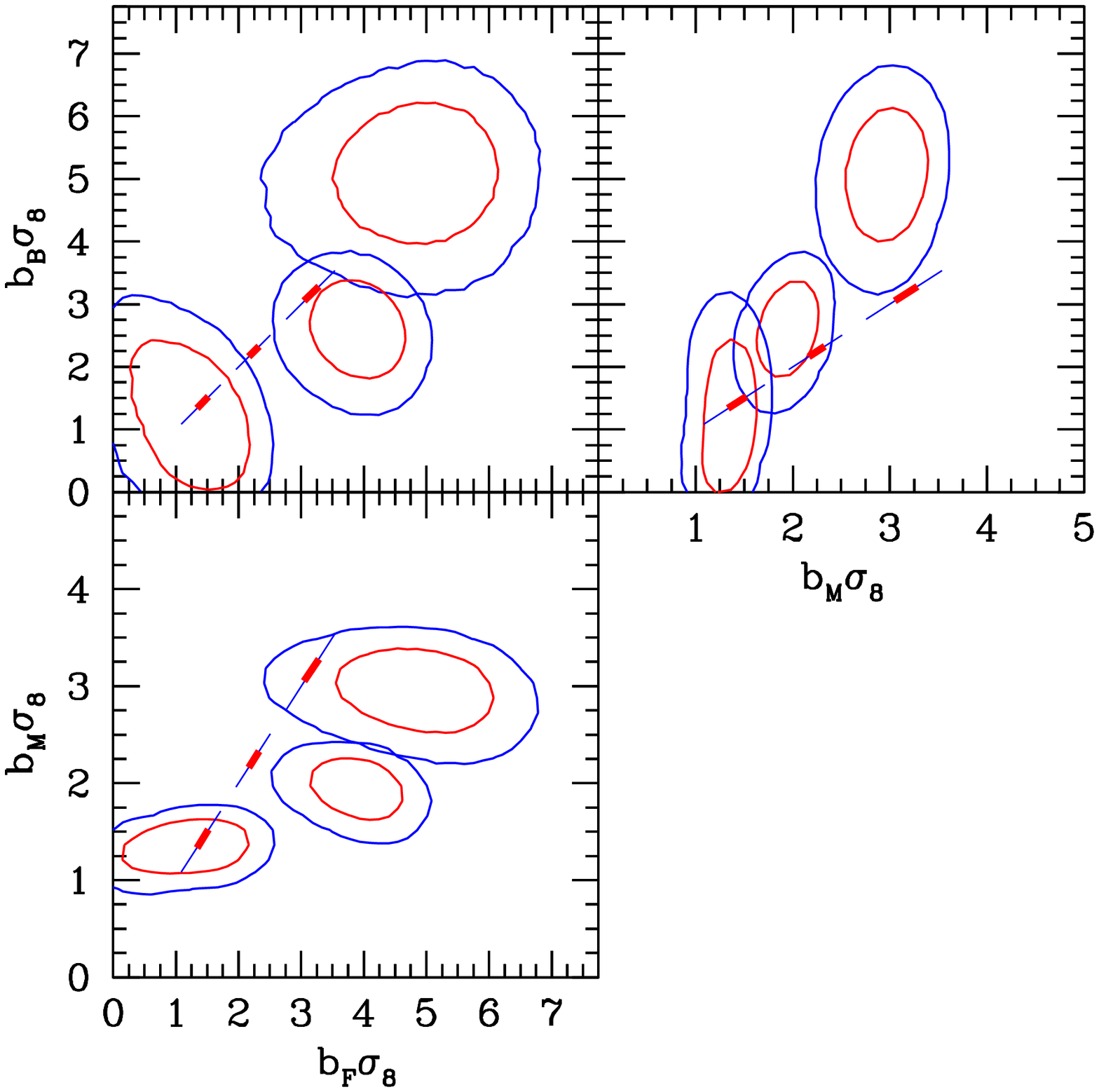,width=0.48\textwidth}}
\caption{{\it Left panel:} 
Results obtained by fitting the data with a three-parameter model (see text) 
where $b_{\rm B}$, $b_{\rm M}$ and $b_{\rm F}$ 
are the bias parameters for the bright, medium and faint 
sub-samples (at fixed redshift), respectively. 
The contours indicate
the 68.3 and 95.4 per cent credibility intervals for the (marginalized) 
joint probability distribution of two bias parameters.
On the diagonal, we indicate the corresponding 
confidence levels for the one-parameter fits. 
The results for the low, medium and high redshif1t samples
are plotted from left to 
right: $b \sigma_8$ clearly increases with 
redshift.  
{\it Right panel:} Same as left panel but using more principal components
of the errors so as to account for 99 per cent of the variance
(as opposed to 95 per cent in the left panel).
Note that errorbars shrink and
$b_{\rm F}$ is shifted by almost 1$\sigma$ with respect to the
left panel.
\label{fig:bias_allcl}}
\end{figure*}

A number of systematic effects can alter the model-fitting procedure
(and thus the statistical significance of the results).
In particular, for strongly correlated data,
the choice of the number of principal components used in the fit
plays a delicate role.
We fixed $N_{\rm pca}$ assuming that our bootstrap errors are accurate
estimates of the real uncertainties and using the $\chi^2$ test. 
Had we decided to include a few more components to account for 99 per cent
of the variance (which is equivalent to consider 7 additional 
components for each redshift sample 
and still gives acceptable values of the reduced $\chi^2$), 
the evidence for clustering segregation would have been stronger. 
In Fig.~\ref{fig:bias_allcl} we compare the solutions for the 
three-parameter models obtained by minimizing $\chi^2_{95}$ (left) and 
$\chi^2_{99}$ (right).
What is shown here are the (marginal) joint probability distributions 
of pairs of bias parameters (for the faint, medium and bright samples). 
Clustering segregation with luminosity is
suggested by the data whenever the contours of the joint distributions
lie away from the diagonal line.
For both $\chi^2_{95}$ and $\chi^2_{99}$, 
this never happens for the low-redshift sample (the bottom-left set of curves)
and the one-parameter solution has to be preferred.
At best, clustering segregation with luminosity is detected at the
$\sim2\sigma$ level (for $\chi^2_{95}$) and at 
$\sim3\sigma$ level (for $\chi^2_{99}$)
in the high- and medium-redshift samples.
Note that, even though a 3$\sigma$ detection is still consistent with pure 
statistical fluctuations, when $\chi^2_{99}$ is used,
all the tests for model selection strongly prefer
either the three- or the six-parameter solution with respect to the
one-parameter fit.
In other words, the results presented in the previous section have to be
considered conservative. 
If one decides to use the full covariance matrix, one will infer
the presence
of a statistically significant clustering segregation with luminosity
in the medium- and high-redshift samples.
Hence, the uncertainty in the number of physical components
is the main limitation of the principal component technique that we used
to account for correlated errorbars.
Bayesian techniques (e.g. Minka 2000) suggest that the components 
which contribute the last few per cent of the bootstrap variance
are most likely dominated by noise. Therefore, we are confident that
the results we presented in section 4.1 are optimal and realistic.
Larger datasets 
with smaller intrinsic uncertainties are thus needed to improve
the significance of our results. 

\begin{figure}
\epsfig{figure=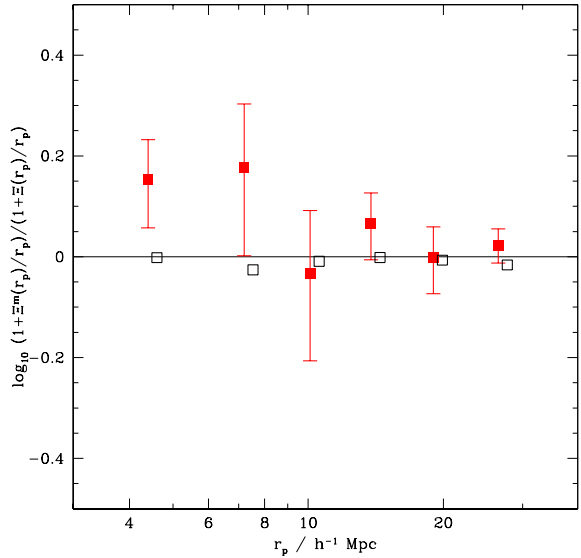,height=8cm}
\caption{Marked correlation function of the medium redshift sample obtained
using the blue luminosity of the quasars as the mark.  
The empty squares (that have been slightly displaced to the right) 
show the mean 
obtained by averaging over one hundred realizations of the same
measurement after randomizing the marks. Errorbars show the standard
deviation around this mean.
If one includes bootstrap errors into the total error budget
(not shown here), the errors on the marked correlation function
typically increase by 20-30 per cent.
\label{fig:marked}}
\end{figure}

In order to cross-check our results with a method affected by different
systematics,
we have computed the marked correlation function of the quasars
(e.g. Sheth, Connolly \& Skibba 2006 and references therein)
using their blue luminosity as the mark.
Results for the medium redshift sample are shown in Fig. \ref{fig:marked}.
Here we
define the marked correlation function (filled squares) as the ratio
of the projected quasar correlation function weighted by the
quasar luminosity (numerator) and the usual projected quasar correlation
function (denominator). By construction, this ratio
is not affected by redshift space distortions.
The open symbols correspond to the mean marked correlation
function obtained by averaging over one hundred realisations 
with randomized marks. To assess
the significance of the marked statistics, 
one usually compares the signal with the standard deviation 
around the mean of the
randomized realisations. From Fig. 4, we see that, on average, the closest
quasar pairs have higher luminosities, thus suggesting the presence, on
those scales, of some clustering segregation. However, given the size of
the errorbars, the statistical significance of this trend is rather
small.
We note that the errors used here only reflect the uncertainty arising
from the distribution of the marks and do not include any uncertainty in
the actual estimate of the correlation function. Considering the corresponding
bootstrap errors (which, for instance, allow for sample variance),
the uncertainties on the marked correlation function typically increase by 
20-30 per cent.
This makes the detection of clustering segregation even more uncertain.
In summary, the analysis of the marked statistics 
reaches exactly the same conclusions as our main work: even though
there is some evidence for luminosity-dependent clustering in our high-redshift
data, our sample is too small to provide a robust detection.
Similar results are obtained for the other redshift bins.

\subsection{Discussion}
When interpreting our results, we have to consider the 
limitations of both the dataset and 
the method used to extract the bias parameters. First, 
the depth of the quasar sample is such that at a given redshift 
we can only probe a factor of ten in luminosity.  
This implies that any luminosity segregation has to be observed 
over this relatively small luminosity range. 
By slicing the 
different redshift bins into six sub-samples,
our analysis tries to extract the maximum possible information from
these limited data.
However, to avoid to be dominated by measurement noise,
we compare narrow luminosity intervals (at both the bright
and faint ends) with larger control samples.
This naturally leads to the selection of partially overlapping luminosity 
ranges. For this 
reason, our samples, despite probing different absolute magnitude 
intervals, are not independent from each other.
Note the striking difference from previous studies of galaxy clustering
(see e.g. Norberg \etal\ 2001, 2002)
where one can perform several independent measurements of the
clustering amplitude as a function of luminosity and span a wide
magnitude range. 

Given these limitations and
based on theoretical models of quasar clustering, should have we expected to 
detect some luminosity dependence in our dataset?
As discussed in the introduction, models make very different predictions.
From Fig. 3 in Lidz et al. (2006) we infer that the models which assume
a tight correlation between quasar luminosity and the mass of the host-halos
predict a difference $\Delta b=(b_{\rm B25}-b_{\rm F25})\,\sigma_8\sim 1$ 
for our high-redshift bin. On the other hand, models based on merger
simulations predict $\Delta b\sim 0.2$. 
Similarly, the semi-analytic models by Kauffman \& Haehnelt (2002) 
predict $0.3\simlt \Delta b \simlt 0.7$ 
depending on the characteristic quasar lifetime (this increases up to
$\Delta b \sim 1$ at $z\simeq 1.5$).
In all cases, the errorbars associated with our measurements are too large to
lead to a statistically significant detection of luminosity dependent
clustering (in fact, our error on $\Delta b$ is $\sigma_{\Delta b}\simeq 2.3$
and we measure $\Delta b\simeq 4.3$ which is only 
$1.4\,\sigma$ and $1.8\,\sigma$ away from the two reference models). 
Assuming that $\Delta b\simeq 1$,
a sample which is nearly fifty times larger than ours (i.e. nearly all-sky) is 
needed to detect the luminosity dependence at the $3\,\sigma$ confidence level
(assuming that the error scales with the sample size, $n$, as $n^{-1/2}$).
Very likely a sample of this size will not be available in the next few years.
Alternatively, one could try to beat the noise by 
reaching a fainter apparent magnitude limit.
Fig. 3 in Lidz et al. (2006) shows that current models predict that, 
by reducing the median luminosity of our F25 sample by a factor of 10, 
one would get $0.6\simlt\Delta b\simlt 2$. 
Note that a value of $\Delta b\sim 2$ could be ruled out at 
the $3\,\sigma$ confidence level by using a sample twelve 
times as large as ours.
In conclusion,
our analysis can only rule out (or detect) extreme luminosity dependent 
biasing 
and future surveys are required to discriminate among current models.
\begin{table*}
\begin{center}
\caption{Best-fitting constant-bias (columns 4,5 and 6) and power-law models 
for six complementary redshift bins. 
The goodness of each fit is measured by the quantity 
$\chi^2_{\rm min}/{\rm dof}$ 
which gives the minimum value assumed by the $\chi^2$ statistic divided
by the number of degrees of freedom.
\label{tab:zdata}}
\begin{tabular}{ccccccccccc}
\hline
\hline
$z_{\rm l}$ & $z_{\rm h}$& $z_{\rm m}$&$b$ & $r_0$&
$\chi^2_{\rm min}/{\rm dof}$ & $r_0$& $\gamma$ &
$\chi^2_{\rm min}/{\rm dof}$ & $r_0^{(\gamma=1.8)}$ &
$\chi^2_{\rm min}/{\rm dof}$\\
& & & & (\Mpc) & &(\Mpc) & & &(\Mpc) & \\
\hline
0.80 & 1.06 & 0.933 &$1.57^{+0.30}_{-0.37}$&$5.5^{+1.5}_{-1.8}$&
$3.47/3$&$4.3^{+2.5}_{-4.2}$&$1.70^{+0.53}_{-0.67}$&$2.02/2$&
$4.9^{+0.8}_{-1.1}$&$2.05/3$\\
1.06 & 1.30 & $1.185$ &$1.76^{+0.35}_{-0.43}$&$5.6^{+1.6}_{-1.9}$&
$1.07/3$&$5.1^{+1.6}_{-4.3}$&
$1.84^{+0.33}_{-0.68}$&$0.87/2$&
$4.9^{+1.0}_{-1.2}$&$0.89/3$\\
1.30 & 1.51 & $1.410$&$2.13^{+0.29}_{-0.33}$&$6.5^{+1.2}_{-1.3}$&
$1.05/3$&$6.1^{+1.2}_{-2.1}$&
$1.93^{+0.44}_{-0.42}$&$1.44/2$&
$5.6^{+0.9}_{-0.9}$&$1.54/3$\\
1.51 & 1.70 & $1.602$&$2.33^{+0.33}_{-0.39}$&$6.7^{+1.2}_{-1.5}$&
$5.33/3$&$4.1^{+2.3}_{-3.9}$&
$1.59^{+0.38}_{-0.55}$&$4.92/2$&
$5.4^{+0.8}_{-0.9}$&$5.17/3$\\
1.70 & 1.89 & $1.796$&$3.02^{+0.45}_{-0.53}$&$8.5^{+1.5}_{-1.9}$&
$0.53/3$&$7.2^{+1.2}_{-2.7}$&
$2.03^{+0.43}_{-0.47}$&$1.19/2$&
$6.3^{+1.0}_{-1.3}$&$1.48/3$\\
1.89 & 2.10 & $1.987$&$4.13^{+0.49}_{-0.55}$&$11.5^{+1.5}_{-1.7}$&
$2.76/3$&$8.8^{+1.2}_{-4.9}$&
$1.82^{+0.27}_{-0.49}$&$2.86/2$&
$8.6^{+1.1}_{-1.2}$&$2.87/3$\\
\hline
\hline
\end{tabular}
\end{center}
\end{table*}

\section{Redshift-dependent clustering}
\label{sec:redshift}

In this section we focus on the redshift dependence of the quasar clustering
amplitude. This is done in real space, using the projected correlation
function. Our study is thus complementary to the redshift-space analysis by 
Croom et al. (2005) which might be affected by large-scale infall.
In this section we also discuss the effect of using different methods for
determining the quasar correlation length.

\begin{figure}
\centerline{\epsfig{figure=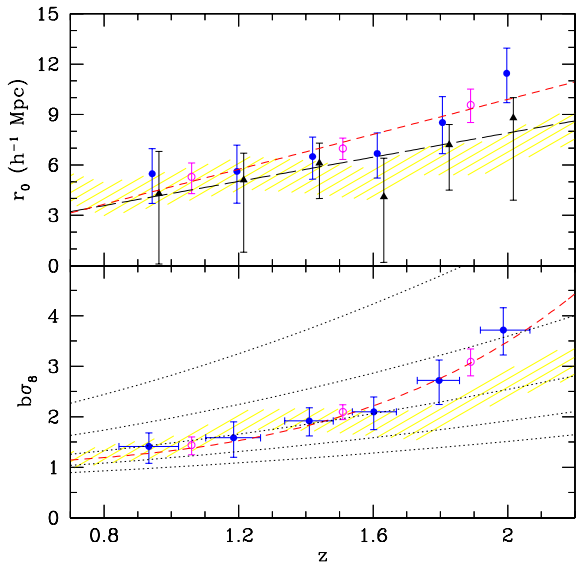,height=8cm}}
\caption{{\it Top:}
Redshift evolution of the quasar correlation length.
Filled circles with errorbars indicate the best-fitting values obtained using
a biased CDM model. 
The short-dashed line is the analytical fit to the data given in equation (\ref{eq:r0fit}).
Filled triangles 
with errorbars refer to the best-fitting values obtained with a 
power-law model (where both $\gamma$ and $r_0$ are free to vary).
These points have been slightly displaced to the right to improve readibility.
The long-dashed line is the linear fit to the data given in the main text.
Empty symbols are analogous to their filled counterparts but
are obtained using larger redshift intervals as in PMN04.  
The shaded area marks the 68.3 confidence levels 
obtained by Croom et al. (2005).
{\it Bottom:} As in the top panel but for the quasar bias parameter.
Horizontal errorbars mark the $16^{\rm th}$ and the $68^{\rm th}$ 
percentiles of the redshift
distribution in a bin.
The dashed line is the fitting function given in equation (\ref{eq:bevo}).
The dotted lines indicate the bias evolution for dark matter halos
with  masses $10^{12},10^{12.5}, 10^{13}, 10^{13.5}$ and $10^{14} M_\odot$ 
(from bottom to top). 
\label{fig:zevo}}
\end{figure}

To improve the analysis performed by PMN04 and better study the evolution with 
cosmic time, 
we split our quasar sample into six redshift bins. 
These are obtained by dividing into two equal sized parts 
(in terms of quasar number)
the redshift bins used in the previous sections and in PNM04.
We first compute the projected auto-correlation functions of all the 
quasars which lie in a given redshift bin.
We then use the method described in \S\ref{sec:data_mod} to
fit these data with the model given in equation (\ref{eq:fit}). 
The best-fitting
models correspond to very good values of the $\chi^2$ statistic, 
indicating that our models accurately describe
the spatial dependence of the measured correlations.
Finally, we 
estimate the correlation length, $r_0$, by determining the scale
at which the best-fitting model has $\xi^{\rm q}(r)=1$.
For comparison, we also fit a power-law model $\xi^{\rm q}(r)=(r/r_0)^\gamma$.
We distinguish two cases where we either
allow both $r_0$ and $\gamma$ to vary or we fix $\gamma=1.8$.
Our results are reported in Table \ref{tab:zdata}
(together with the corresponding 68.3 per cent confidence levels) and
plotted in Fig. \ref{fig:zevo}. 
Note that the best-fitting values for the quasar correlation length derived 
from the power-law model are systematically lower than the CDM ones (the two 
estimates are anyway consistent within $1\sigma$ uncertainties). 
This is because, in a CDM model, the power-law index of the mass 
auto-correlation function is a function of the spatial separation.
For instance, $\gamma_{\rm eff}=d\ln \xi/d\ln r\simeq 1.3$ at $r=5$ \Mpc, 
while $\gamma_{\rm eff}\simeq 1.9$
at $r=20$ \Mpc. Similarly, the projected correlation function, 
$\Xi(r_\perp)/r_\perp$, has a slope varying between $1.6$ and 2.3
when $5<r<20$ \Mpc. 
Therefore, a CDM model has a much lower correlation on large-scales with
respect to a power-law model with the same $r_0$ (and $\gamma<2$). 
It is then clear from equation (\ref{eq:xibar}) that
a CDM model needs a higher overall normalization than a power-law
model to fit the same projected correlation. 

The evolution of the quasar 
correlation length between $0.8<z<2.1$ can be approximately described
by the linear relation
\footnote{A nearly perfect interpolant between our measured points is 
$r_0=\left[5+\left(\frac{1+z}{2.4}\right)^{8}\right] h^{-1}\,{\rm Mpc}$.
This, however, ignores the statistical uncertainties in the measure of $r_0$
(i.e. has $\chi^2\ll 1$).}
\begin{equation}
r_0=\left[7.3+5.2\,(z-1.5)\right]  h^{-1}\,{\rm Mpc}
\label{eq:r0fit}
\end{equation}
(this becomes $r_0=[6.1+3.6\,(z-1.5)]$ \Mpc \ for the values
of $r_0$ determined with the power-law fit to $\Xi(r_\perp)/r_\perp$).
For $0.8<z<1.6$, our results are in good agreement with Croom et al. (2005),
but we find evidence for a stronger variation at high-z 
(see Fig. \ref{fig:zevo}). 
\footnote{Our estimates of the uncertainties for $r_0$ and $\gamma$ are
nearly a factor of 2 larger than in Croom et al. (2005). This 
is due to a number of facts. On one hand, we find 
that blockwise bootstrap errors on the redshift-space correlation 
function are typically 30\% smaller than on the projected correlation function.
On the other hand,
Croom et al. (2005) assume statistically independent Poisson error 
bars for the redshift-space correlation function at separations smaller than 
50 \Mpc. In particular, neglecting correlations between points at different
spatial separations results into smaller errors for the fitted parameters.}
This is more evident when we use the CDM model to fit the data.
The models by Kauffman \& Haehnelt (2002) with a quasar lifetime of $\sim 10^7$ 
yr match our data at $z<1.7$. However, as already pointed out in PMN04,
an increase in the lifetime by nearly an order of magnitude is required
to reproduce the biased-CDM fits at $z\sim 2$.

Given that the mass autocorrelation function rapidly increases with cosmic
time, the bias parameter of our quasars has to increase with $z$.
The bottom panel of Fig. \ref{fig:zevo} shows the evolution of $b$.
Note that, within the range of cosmological models allowed by observations, the
quasar bias parameter scales linearly with $\sigma_8$.
For this reason we decided to plot the product 
$\sigma_8^{\rm q}=b\,\sigma_8$ which is
independent on the assumed value for the amplitude of the linear power 
spectrum.   
Similarly to the correlation length, $\sigma_8^{\rm q}$ seems to increase
rapidly for $z>1.6$ and 
the bias evolution is well approximated by the relation
\begin{equation}
b\sigma_8=1+\left(\frac{1+z}{2.5}\right)^{5}\;.
\label{eq:bevo}
\end{equation}
In CDM models,
the clustering amplitude of
massive dark-matter halos mainly depends on their mass. 
It is then interesting to see what halo masses correspond to the bias 
parameters of the 2QZ quasars. 
In the bottom panel of Fig. \ref{fig:zevo}, we plot the evolution of 
$\sigma_8^q$ for a number halo masses which is obtained using
the model by Sheth \& Tormen (1999).
\footnote{
A concordance cosmology with $\sigma_8=0.9$ is assumed here 
but results depend only slightly on $\sigma_8$.}
The observed bias evolution is
well reproduced by assuming that quasars are associated with halos of mass
$\sim 10^{13} M_\odot$.
This is fully consistent with the more detailed analysis performed in PMN04.

\section{Summary}
\label{sec:discussion}

We have used a flux limited sample of $\sim 14,000$ 
2QZ quasars brighter than 
M$_{\rm b_{\rm J}}-5\log_{10}h=-21.7$ to study the 
quasar clustering properties in the redshift range 
$0.8<z<2.1$. Our main results are summarized as follows.

(i) Splitting the sample in three redshift intervals each divided into
six luminosity ranges and 
combining information from the corresponding
auto- and cross-correlation functions,
we find some evidence for clustering segregation with luminosity.
For redshifts $z>1.3$,
a frequentist model selection technique (the F-test) prefers a multi-parameter
fit to the data at the 95 per cent confidence level 

(ii) A number of statistical 
tests based on information theory and Bayesian techniques show
weak evidence for luminosity dependent clustering at high redshift 
($z>1.3$) and no evidence at low redshift ($z<1.3$).
These results somewhat depend on the number of principal components
used in the fitting procedure. Accounting for a larger fraction of the
bootstrap variance increases the significance of the detection of
clustering segregation with luminosity. 

(iii) 
Larger datasets, possibly with a deeper coverage,
are needed to discriminate among current models of quasar formation
and to pin-point the detailed quasar clustering trends as a function of 
luminosity at a given redshift.

(iv)
Splitting the sample into six complementary redshift bins, we find strong
evidence for an increase of the clustering amplitude with lookback time.
We detect pure quasar-clustering evolution between 
$z_{\rm eff}=0.93$ and $z_{\rm eff}=1.99$ at the $2.7\sigma$ 
confidence level.
A linear fit for the evolution of $r_0$ with redshift 
is given in equation (\ref{eq:r0fit}).

(v)
Accounting for the evolution of the mass density in a $\Lambda$CDM model,
we find that the high-redshift quasars 
($z_{\rm eff}=1.99$) are $\sim 2.6$ times more biased 
than their low-redshift counterparts ($z_{\rm eff}=0.93$).
Evolution in $b$ is detected at the $4.3\sigma$ confidence level.

(vi) The clustering amplitude of optically selected quasars suggests that
they are hosted by halos with mass $M\sim 10^{13} M_\odot$ (see also
PMN04).

\section*{ACKNOWLEDGMENTS}

PN acknowledges support from the Zwicky Prize fellowship program at
ETH-Z\"urich and the PPARC PDRA fellowship held at the IfA.
We thank Martin White, Andrey Kravtsov and Andrew Zentner for useful 
discussions.
The 2dF QSO Redshift Survey (2QZ) was compiled by the 2QZ survey team
from observations made with the 2-degree Field on the Anglo-Australian
Telescope.

\appendix
\section{Model selection criteria}
{\it The F-test.} 
If the uncertainties of the measurements are known to a good precision,
model selection can be performed using a simple $\chi^2$ test.
However, we cannot prove that 
our bootstrap errors accurately reproduce the true scatter in 
the data. 
An alternate 
test that does not require knowledge of the true standard 
deviation (up to a constant scaling factor) is to form the ratio
\begin{equation}
{\cal F}=\frac{\chi^2_{n-k}-\chi^2_{n}}{k}\frac{N-n}{\chi^2_{n}}
\end{equation}
where $N$ is the number of independent data values used in the 
fit (in our case $N=N_{\rm pca}$).
The numerator in this equation is the difference between 
the $\chi^2$ calculated for $n$ parameters and that for 
$n-k$ parameters.
Being a ratio of two $\chi^2$-distributed variables, ${\cal F}$ 
follows an $F$-distribution with $(k,N-n)$ degrees of freedom.
We can then estimate how significant is the addition of $k$ 
parameters by integrating this distribution from 0 to ${\cal F}$.

\noindent
{\it The Akaike information criterion.}
The question to find which model best approximate a given dataset can
be addressed in terms of information theory:
the best model minimizes the loss of information.
The Akaike information criterion (AIC)
evaluates models using the Kullback-Leibler information (Akaike 1973).
In terms of the number of independent datapoints, $n$, and
of model parameters, $p$, 
\footnote{In our case $p$ is the number of model parameters plus
one to account for the estimation of the $\chi^2$ function.}
\begin{equation}
{\rm AIC}=\chi^2_{\rm min}+2p+\frac{2p\,(p+1)}{n-p-1}
\end{equation}
where the second order term on the right is needed when
the size of the dataset does not
exceed the number of model parameters by a large factor ($\simgt40$).
The model with the minimum AIC has to be considered the best.
Note that
the AIC penalizes for the addition of parameters and thus selects a model
that fits well but has a minimum number of parameters.
The relative probability that a model is the correct solution is given
by the Akaike weights
\begin{equation}
w_i=\frac{e^{-\Delta{\rm AIC}_i/2}}{\sum_{k=1}^N e^{-\Delta{\rm AIC}_k/2}}
\end{equation}
where the indices $i$ and $k$ run over the different models and
$\Delta{\rm AIC}={\rm AIC}_i-\min_k({\rm AIC}_k)$.
In general, models receiving AIC within 2 of the ``best'' deserve
consideration, those within 3-7 have considerably less support and
those above 7 are basically rejected by the test (Burnham \& Anderson 1998).

\noindent
{\it The Bayes factor.}
Bayes factors are the dominant method for Bayesian model testing
(see Kass \& Raftery 1994 for an extensive review).
The Bayes factor, $B_{12}$, is the ratio between the marginalized likelihoods
(i.e. between the probabilities of the data given the models)
for two different models (1 and 2) and provides
a scale of evidence in favour of a model versus another.
The model with the highest marginalized likelihood is the best. 
Following an early suggestion by Jeffreys (1961),
Kass \& Raftery (1994) proposed the following
``rule of thumb'' for interpreting the Bayes factors:
$1<B_{12}<3$ provides weak evidence (barely worth mentioning) for model 1,
$3<B_{12}<12$ provides substantial evidence for model 1,
$12<B_{12}<150$ provides strong evidence for model 1,
$B_{12}>150$ provides decisive evidence for model 1. 
Several numerical approaches have been proposed to compute Bayes factors
based on MCMC sampling 
but most of these methods are subject to numerical or stability
problems (see e.g. Han \& Carling 2000 for a review).
As a compromise between accuracy and coding complexity 
we estimate  the Bayes factors using the harmonic mean estimator
(see e.g. Kass \& Raftery 1994).
We repeat the calculation using three different chains and we use the
standard deviation among them as an estimate of the uncertainty 
of our marginalized likelihoods.

\noindent
{\it The Bayesian information criterion.}
Unfortunately, while Bayes Factors are rather intuitive, as a matter
of fact they are often quite difficult to calculate.
This makes simpler (but approximate) estimates of the Bayes factors
of great interest.
Schwarz (1978) derived the Bayesian information criterion (BIC) 
as a large sample approximation to twice the logarithm of the Bayes
factor.
\begin{equation}
{\rm BIC}_1-{\rm BIC}_2\simeq -2 \ln (B_{12})\;,
\end{equation}
with
\begin{equation}
{\rm BIC}=\chi^2_{\rm min}+p\,\ln n\;.
\end{equation}
Similarly to the AIC,
the preferred model is the one with the lowest value of the criterion.
The Kass-Raftery criterion for model selection is also applied to the BIC.

\noindent
{\it The Deviance Information Criterion.}
Spiegelhalter et al. (2002) have recently proposed 
a Bayesian generalization of the AIC:
the deviance information criterion (DIC).
It is based on the posterior distibution of the log-likelihood 
or deviance ($D=-2\log {\cal L}$ which, in our case, coincides with $\chi^2$
function):
\begin{equation}
{\rm DIC}=\langle D \rangle+p_D
\end{equation}
where $\langle D \rangle$ is the posterior expectation of the deviance 
while the effective number of parameters, $p_D$, is 
defined as the difference between the posterior mean of the
deviance and the deviance evaluated at the posterior mean of the parameters.
$p_D=\langle D \rangle-D(\langle \theta \rangle)$.
The DIC can be re-written as
\begin{equation}
{\rm DIC}=D(\langle \theta \rangle)+2p_D
\end{equation}
which makes the analogy with the first-order AIC explicit.
A good model corresponds to a low DIC and
model-selection criteria developed for the AIC appear to work well also for
the DIC (Spiegelhalter et al. 2002).
Note, however, that,
since $p_D\leq p$, the DIC tends to be less conservative than the AIC
in terms of model complexity.
The main attraction of using this measure is that it is trivial to compute
when performing MCMC on the models.

\end{document}